\documentclass[aip,apl,superscriptaddress,amsmath,amssymb,graphicx, floatfix
, reprint
]{revtex4-1}
\usepackage{dcolumn}
\usepackage{bm}
\usepackage{mathptmx}
\usepackage[english]{babel}
\usepackage[utf8]{inputenc}
\usepackage{verbatim}
\usepackage[separate-uncertainty=true]{siunitx}
\usepackage{booktabs}
\usepackage[version=3]{mhchem}
\usepackage[export]{adjustbox}
\usepackage[caption=false]{subfig}
\usepackage{graphicx}
\usepackage[capitalise]{cleveref}

\usepackage{color}
\graphicspath{ {./Figures/} }

\newcommand{\stabheight}{6cm}

\begin{document}

\title{Observation of single phonon-mediated quantum transport in a silicon single-electron CMOS transistor by RMS noise analysis}

\author{Stefano Bigoni}
\affiliation{Istituto di Fotonica e Nanotecnologie, Piazza Leonardo da Vinci 32, 20133 Milan, Italy}
\author{Marco L. V. Tagliaferri}
\affiliation{Istituto di Fotonica e Nanotecnologie, Piazza Leonardo da Vinci 32, 20133 Milan, Italy}
\affiliation{Université Grenoble Alpes, CEA, IRIG-Pheliqs, 38000 Grenoble, France}
\author{Dario Tamascelli}
\affiliation{Università degli Studi di Milano, Dip. di Fisica ``Aldo Pontremoli'', Via Celoria 16, I-20133 Milano, Italy}
\author{Sebastiano Strangio}
%\affiliation{}
\author{Roberto Bez}
\author{Paolo Organtini}
\affiliation{LFoundry, Via A. Pacinotti 7, 67051 Avezzano, L'Aquila, Italy}
\author{Giorgio Ferrari}
\affiliation{Politecnico di Milano, Piazza Leonardo da Vinci 32, 20133 Milan, Italy}
\author{Enrico Prati}
\email{enrico.prati@cnr.it}
\affiliation{Istituto di Fotonica e Nanotecnologie, Piazza Leonardo da Vinci 32, 20133 Milan, Italy}

\date{\today}

\begin{abstract}
We explore phonon-mediated quantum transport through electronic noise characterization of a commercial CMOS transistor. The device behaves as a single electron transistor thanks to a single impurity atom in the channel. A low noise cryogenic CMOS transimpedance amplifier is exploited to perform low-frequency noise characterization down to the single electron, single donor and single phonon regime simultaneously, not otherwise visible through standard stability diagrams. 
Single electron tunneling as well as phonon-mediated features emerges in rms-noise measurements. Phonons are emitted at high frequency by generation-recombination phenomena by the impurity atom.
The phonon decay is correlated to a Lorentzian $1/f^2$ noise at low frequency.
\end{abstract}

\maketitle

%%%%%%%%%%%%%%%%%%%%%%%%%%%%%%%%%%%%% Introduction %%%%%%%%%%%%%%%%%%%%%%%%%%%%%%%%%%%%%%%

%Objective of the work and why it is important (starting sentence)
Single-phonon mediated quantum transport is observed through electronic noise characterization in a single electron quantum dot induced in an industrial silicon transistor operated at cryogenic temperature.
%<Background>
    %                What else has been done?
Phonon-mediated quantum transport has been reported as a byproduct of Coulomb blockade electronic spectroscopy of semiconductor quantum dots at cryogenic temperatures\cite{fujisawa1998spontaneous,zwanenburg2009spin,Escott10,granger2012quantum,braakman2013photon,braig2003vibrational}.
%based on the amplification of tiny impedance  mismatch between a microwave line and the few electron system confined by a quantum dot. The need of removing disturbances has led in the past to develop cryogenic electronics, which is specially beneficial for quantum information processing on some qubit platforms such as semiconductor and superconductors.\cite{CYO ELECTRONICS}
%                How?
On the other hand, the need of spectroscopy methods for the characterization of quantum dots in the few-electron (or hole) regime has led to reflectometry \cite{crippa2019gate} which probes charge tunneling transitions through the dispersive shift of a resonator connected to a gate electrode.

%Different technologies have been adopted ranging from Heterojunction-Bipolar-Transistor\cite{tracy2016single} to cryogenic transimpedance amplifier built with CMOS technologies. 
%In particular, being spin-qubits states typically read out via charge sensors, cryogenic current amplifiers have been introduced\cite{tagliaferri16,tracy2016single}.

    %                What have WE done previously?
We have already demonstrated the control of industrial and pre-industrial silicon nanotransistors at cryogenic temperatures, down to the single electron\cite{turchetti15}%[Enrico-few electron]
and single atom regime\cite{leti2011switching,crippa2015valley}, respectively. %[Crippa Kondo].
We developed custom cryogenic CMOS amplifiers \cite{Guagliardo2013} to study single charge dynamics 
%\cite{mazzeo12} 
and reported performances \cite{tagliaferri16} compatible with the readout of spin qubits in silicon\cite{prati2013quantum}.      
    %        <Objectives of the work >
Here we explore the microscopic nature of phonon-mediated quantum transport through the current of an industrial CMOS silicon transistor at 4.2 K using a cryogenic CMOS transimpedance amplifier. 
In the past, noise characteristics in nanowires have already been explored\cite{vitusevich2017noise}, revealing insights on the microscopic nature of the processes, including generation-recombination phenomena.
Here, we base our study on the characterization of the electronic noise instead of current as usual. We exploit it to magnify the single atom regime, where few phonons processes occur. 
Indeed, because of their relatively small magnitude, phonon-mediated quantum transport becomes visible only by Root Mean Square (RMS) noise measurements.
   
    %        <Guidance to the reader>
    %               What should the reader watch for in the paper?
    %                What are the interesting high points?
    %                What strategy did we use?
By exploiting a cryogenic CMOS transimpedance amplifier, we identify additional quantum transport features generated by few phonon-emission at the single impurity site and we observe a low frequency noise spectral density behaving as $1/f^2$.
%We measure both dc and RMS noise current, as well as low-frequency noise spectra.   
    %        <Summary/Conclusion>
    %                What should the reader expect as a conclusion?
More in general, we show that noise measurements provide complementary knowledge with respect of standard DC current characterization, thanks to the information provided by both the RMS and the spectrum of the noise of the current.
%%%%%%%%%%%%%%%%%%%%%%%%%%%%%%%%%%%%%%%%%%%%%%%%%%%%%%%%%%%%%%%%%%%%%%%%%%%%%%%%%%%%%%%%%%%%%%

%%%\subsection{Device and Methods}
        
          \begin{figure*}[htpb]
                \centering
                    \subfloat[]{\label{fig:setup}
                        \includegraphics[width=0.24\textwidth]{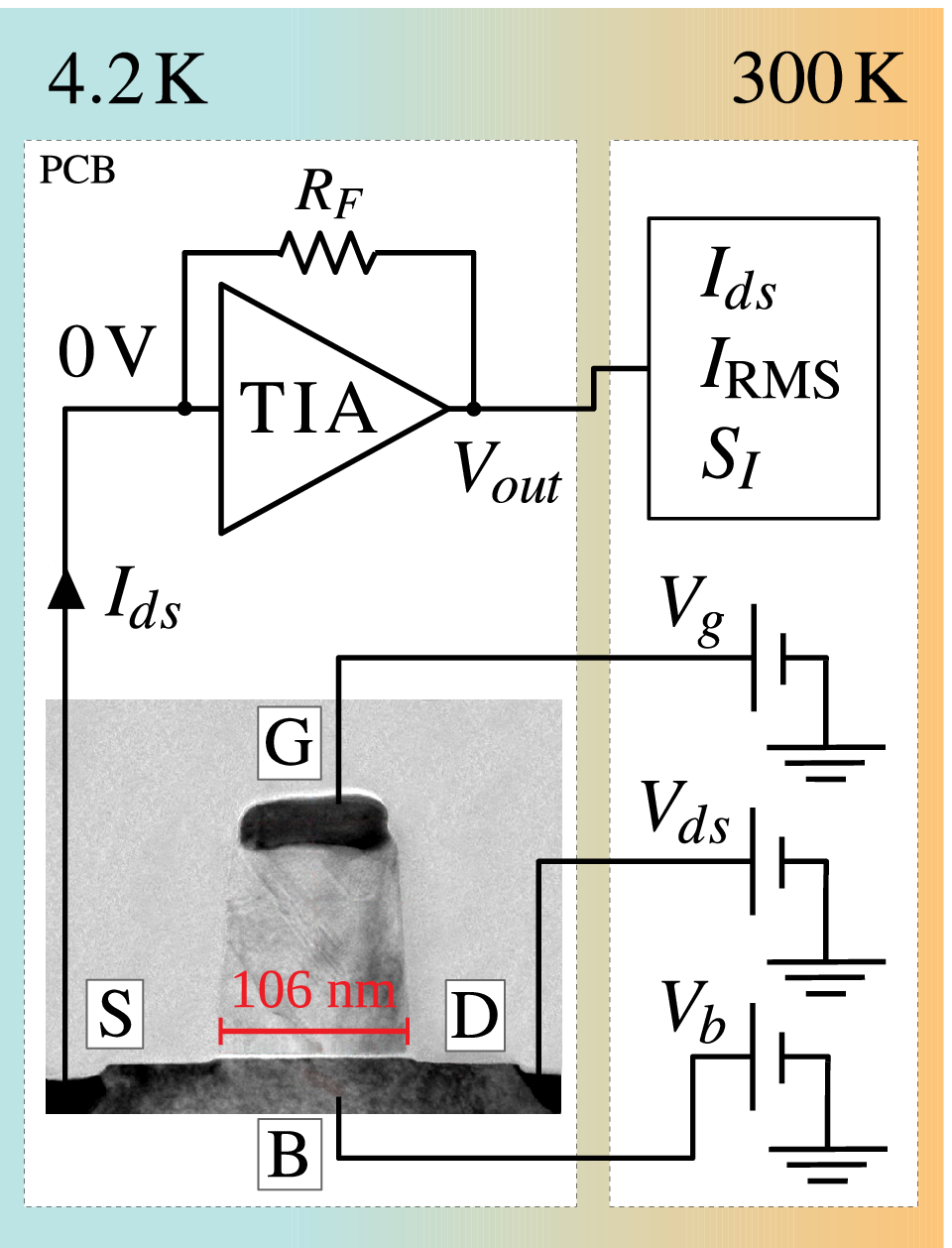}}
                    \subfloat[]{\label{fig:ivtrace}
                        \includegraphics[width=0.33\textwidth]{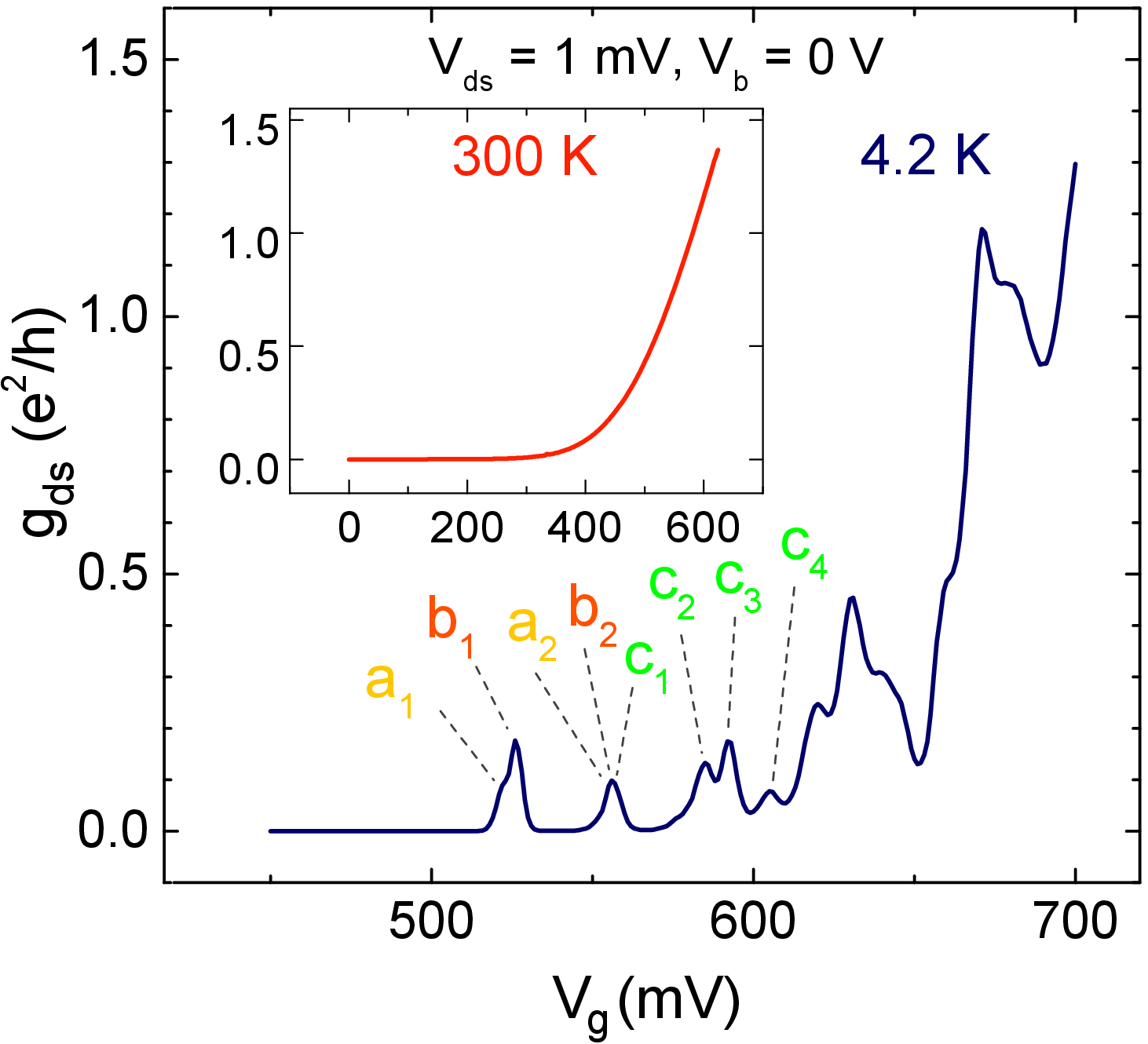}} \;
                    \subfloat[]{\label{fig:bulk}             
                        \includegraphics[width=0.4\textwidth]{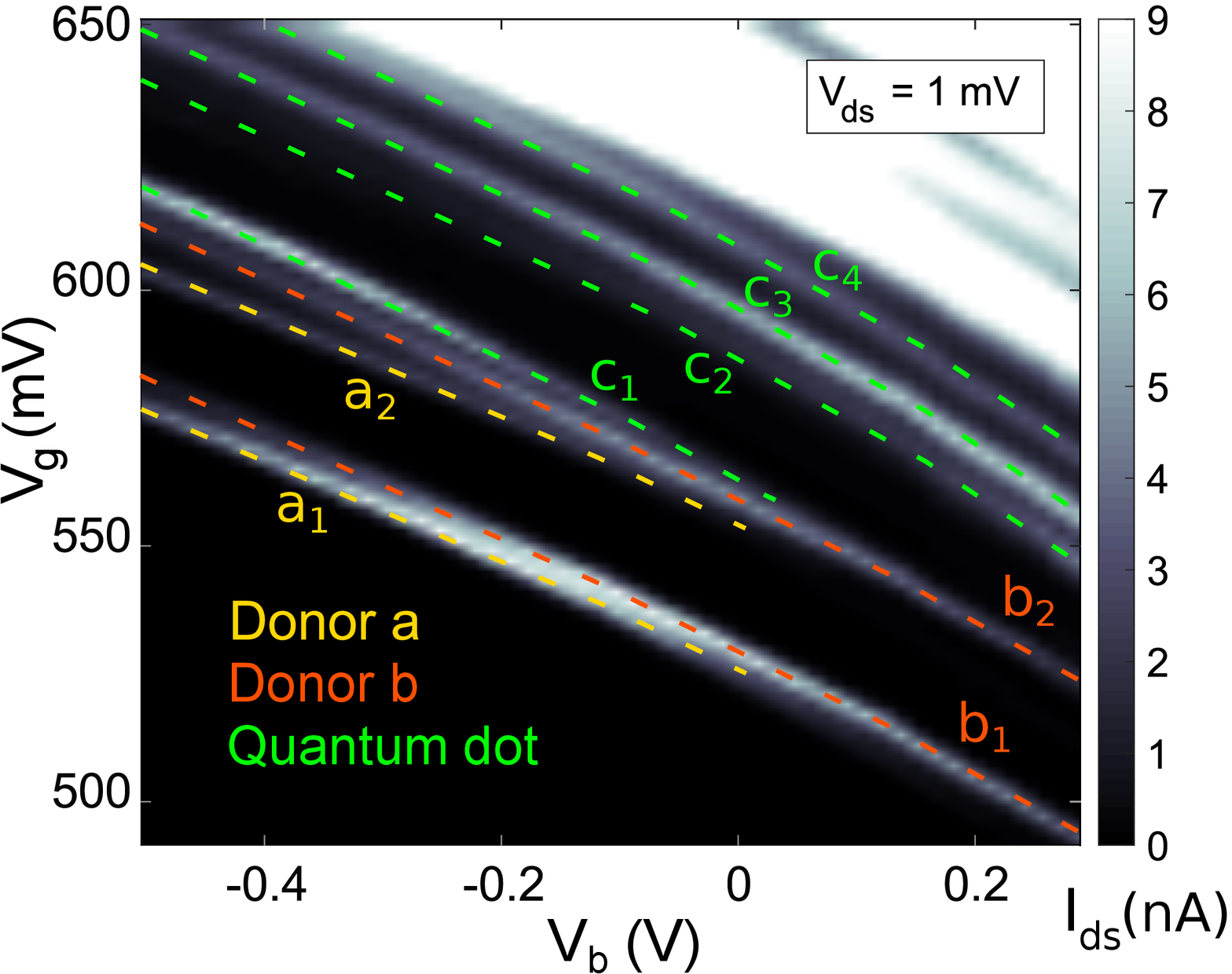}}
                   \caption{
                            \protect\subref{fig:setup} TEM image of a device nominally identical to the device under test, with basic schematic of the measurement setup. The main element of the measurement circuit consists of the cryogenic transimpedance amplifier (TIA) operating at the same temperature of the device. In red, it is reported the actual channel length of 106 nm.
                        \protect\subref{fig:ivtrace} Device transconductance ($g_{ds}=I_{ds}/V_g$) at both \SI{4.2}{\kelvin} and room temperature, respectively: Coulomb blockade (CB) peaks are visible at \SI{4.2}{\kelvin}, in contrast to the standard n-type MOSFET behavior at room temperature. The peaks associated to the donor states exhibit high conductance in terms of $e^2/h$ indicating a strong coupling regime.
                        CB peaks are labeled as in the $V_g-V_b$ stability  diagram.
                            \protect\subref{fig:bulk} Source-drain current $I_{ds}$ versus both the gate and the bulk voltage. The bulk electrode is operated as a back-gate: the CB peaks at different bulk voltage form parallel lines which give information about the localized states. 
                        Different slopes are associated to distinct islands: more specifically the red (b$_1$, b$_2$) and the yellow (a$_1$, a$_2$) dashed lines correspond to two donors, while the green ones (c$_1$ to c$_4$) are associated to a disorder related electrostatically defined quantum dot.
                    }
                   
            \end{figure*}

        The single-gated silicon n-type MOSFET is produced by LFoundry by a standard \SI{110}{nm} CMOS lithography process, with a nominal supply voltage of \SI{1.2}{\volt}.
        Although several different samples with different form factors $W\times L$ have been characterized, for the sake of consistency we report the characterization of the most representative one, of nominal width $W=\SI{140}{\nano\meter}$ and length $L=\SI{100}{\nano\meter}$. 
        %This device presents four electrodes, as shown in \cref{fig:setup}: source, drain, gate and bulk,  A bulk contact is present which is used as a back-gate.
        The source electrode is kept at zero potential and, in the following, potentials applied to the other terminals are referred to it.
        % At room temperature the device behaves as a standard n-type MOSFET, its threshold voltage is estimated from a linear fit of the points with highest slope in the inset of \cref{fig:ivtrace} as $V_{th} = \SI{0.471}{\volt}$.
        %(The 50 nA limit in measurements may not be enough for measuring the real maximum transconductance and then the real threshold voltage)
        In this kind of devices, below \SI{10}{K}, disorder at the \ce{Si}/\ce{SiO2} interface \cite{mazzeo12,tagliaferri2016compact} dominates electron confinement, as along as the effect of a gate voltage $V_g$ near threshold gives rise to local electrostatic potential minima through which electron tunneling can occur.
        As a consequence, the formation of one or more quantum dots in the channel of the transistor arises.
        In addition, in short channel devices (below 100-200 nm) donor atoms (here As) can be randomly diffused from the source or drain contacts in the channel region, sufficiently far to lie approximately in the center of the channel and thus contributing to additional quantum transport channels\cite{leti2011switching}.  %GRUPPO REFS 2
        
        Measurements were carried out at \SI{4.2}{\kelvin}, by dipping the samples in liquid \ce{^4He}, in order to observe electron localization and quantum transport in the device.
        We used a custom-made steel dip-stick provided of a 16-pin dual-in-line carrier and a custom cryogenic amplifier\cite{mazzeo12, tagliaferri16}, described below, as a pre-amplification stage.
        Such cryogenic amplifier allowed to obtain simultaneous measurements of source-drain DC current $I_{ds}$ and its rms noise level $I_{\mathrm{RMS}}$, or alternatively, coupled to a custom spectrum analyzer\cite{sampietro1999}, the noise power spectral density $S_I$ at constant applied voltages (Figure 1a).
        Other relevant quantities like the differential conductance $g_{ds,\mathrm{diff}}={\delta I_{ds}}/{\delta V_{ds}}$ and transconductance ${g_{m,\mathrm{diff}} = \delta I_{ds}}/{\delta V_{g}}$ are calculated from the DC measurements.
        
        %%%\subsection{Cryogenic amplifier}
        The key element of our measurement setup is a custom CMOS amplifier designed to operate at \SI{4.2}{\kelvin}, which reduces the thermal noise of the resistors and transistors.
        In addition, the length of the connections between the cryogenic device and the first amplification stage is reduced from meters of a standard room temperature solution to the scale of the centimeter, thus cutting most of the parasitic capacitance and correspondingly the high frequency noise on current measurements, with respect to room-temperature amplification.
        The amplifier uses a transimpedance architecture with a resistive feedback that sets the current-to-voltage conversion factor. The feedback resistor $R_F$ is chosen as a trade-off between the input-referred current noise and the signal bandwidth, both inversely proportional to the resistor value. 
        %For more flexibility, the cryogenic amplifier has a digital signal to select two different feedback resistors, i.e. an on-chip resistor of \SI{25}{\mega\ohm} at \SI{4.2}{\kelvin} providing an amplifier bandwidth of about \SI{20}{\kilo\hertz} or, alternatively, an off-chip resistor of \SI{200}{\mega\ohm} at 4.2 K and bandwidth of \SI{2.8}{\kilo\hertz}. 
        For better consistency, all the measurements reported in this work are taken with a feedback resistor of \SI{200}{\mega\ohm} at 4.2 K that provides a bandwidth of \SI{2.8}{\kilo\hertz} and a noise level below to \SI{10}{\femto\ampere\per\sqrt{\hertz}}. The cryogenic amplifier has been implemented in a standard \SI{0.35}{\micro\metre} CMOS technology previously characterized at \SI{4.2}{\kelvin} \cite{Guagliardo2013}, it occupies \SI{0.3}{\milli\metre^2} and has a static current consumption of \SI{600}{\micro\ampere} with a power supply of \SI{3.3}{\volt}.
        
\begin{figure*}[htbp]
               \centering
                   
                        \subfloat[]{
                            \includegraphics[height=\stabheight]{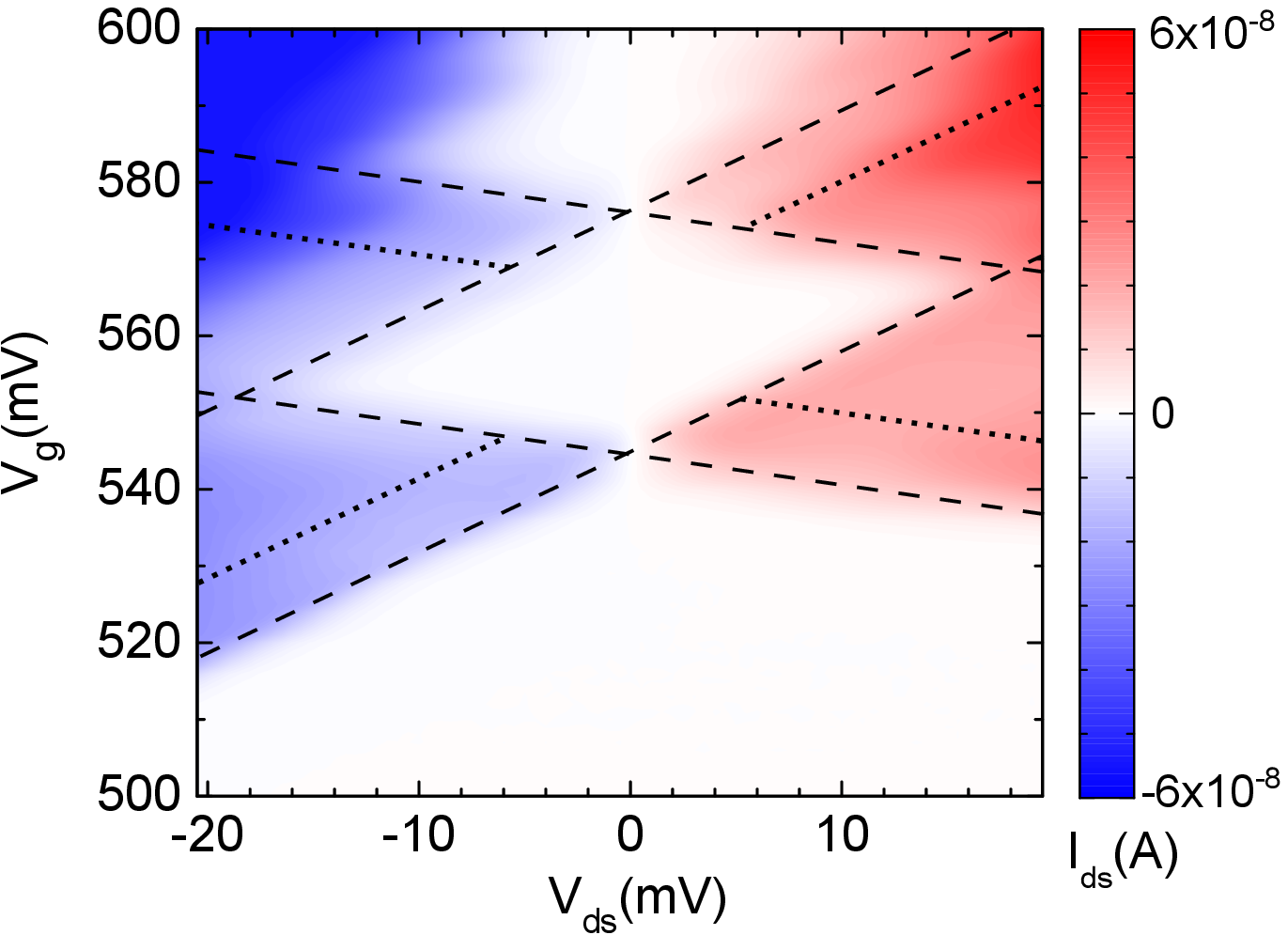}
                            \label{fig:stab-curr}
                            }
                        \subfloat[]{
                            \includegraphics[height=\stabheight]{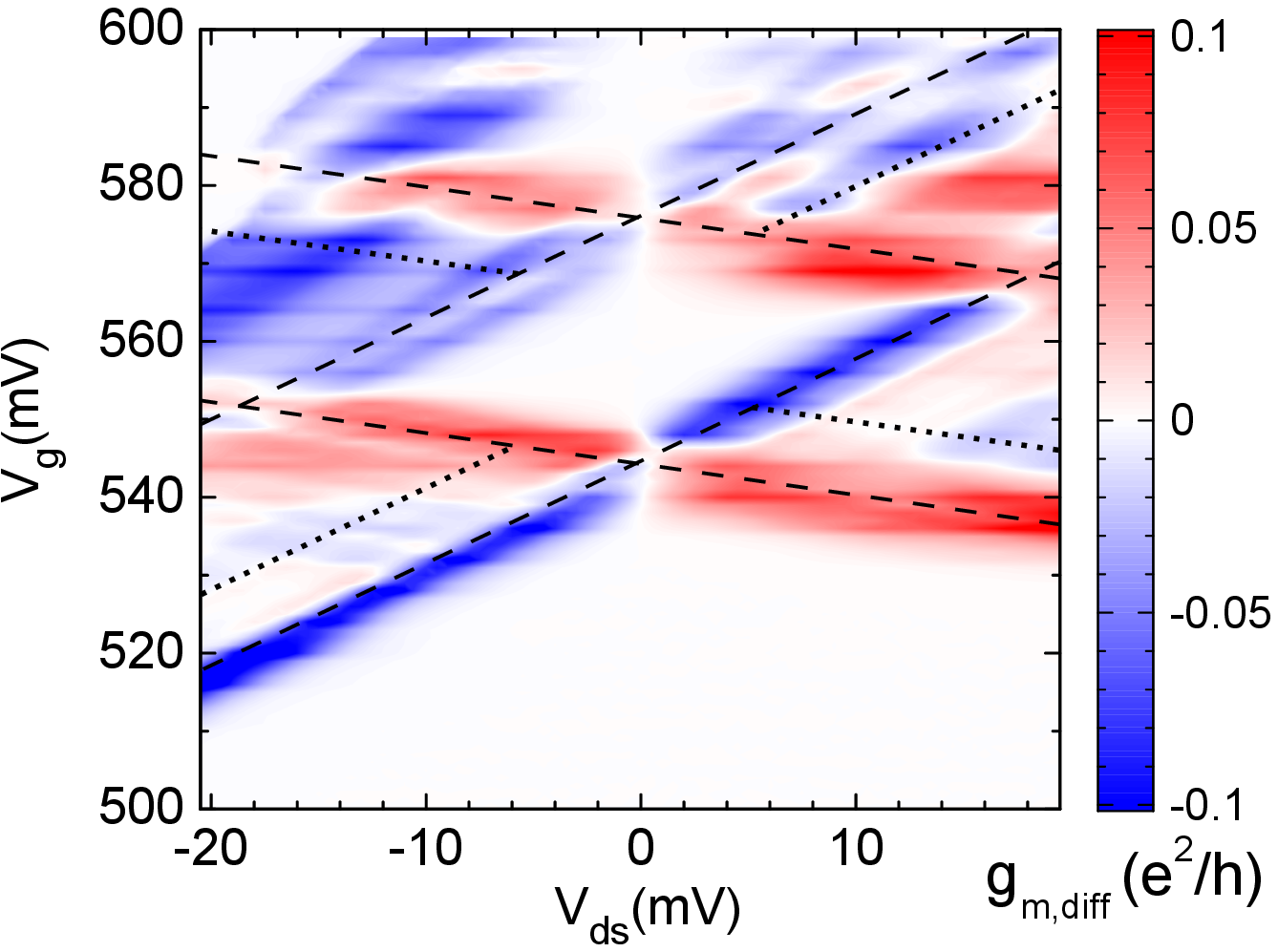}
                            \label{fig:stab-condg}
                            } \\
                        \subfloat[]{
                            \includegraphics[height=\stabheight]{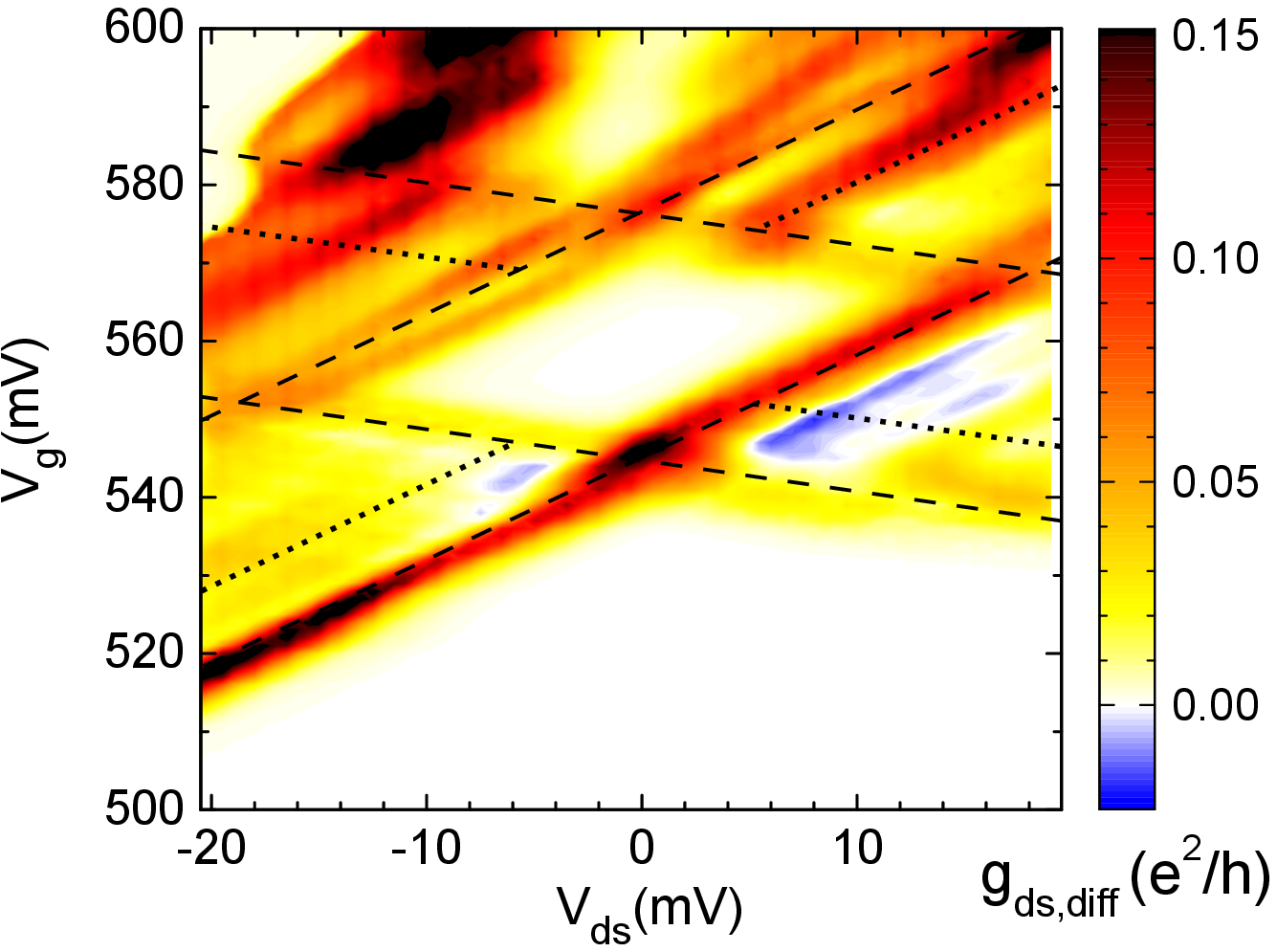}
                            \label{fig:stab-condd}
                            }
                        \subfloat[]{
                            \includegraphics[height=\stabheight]{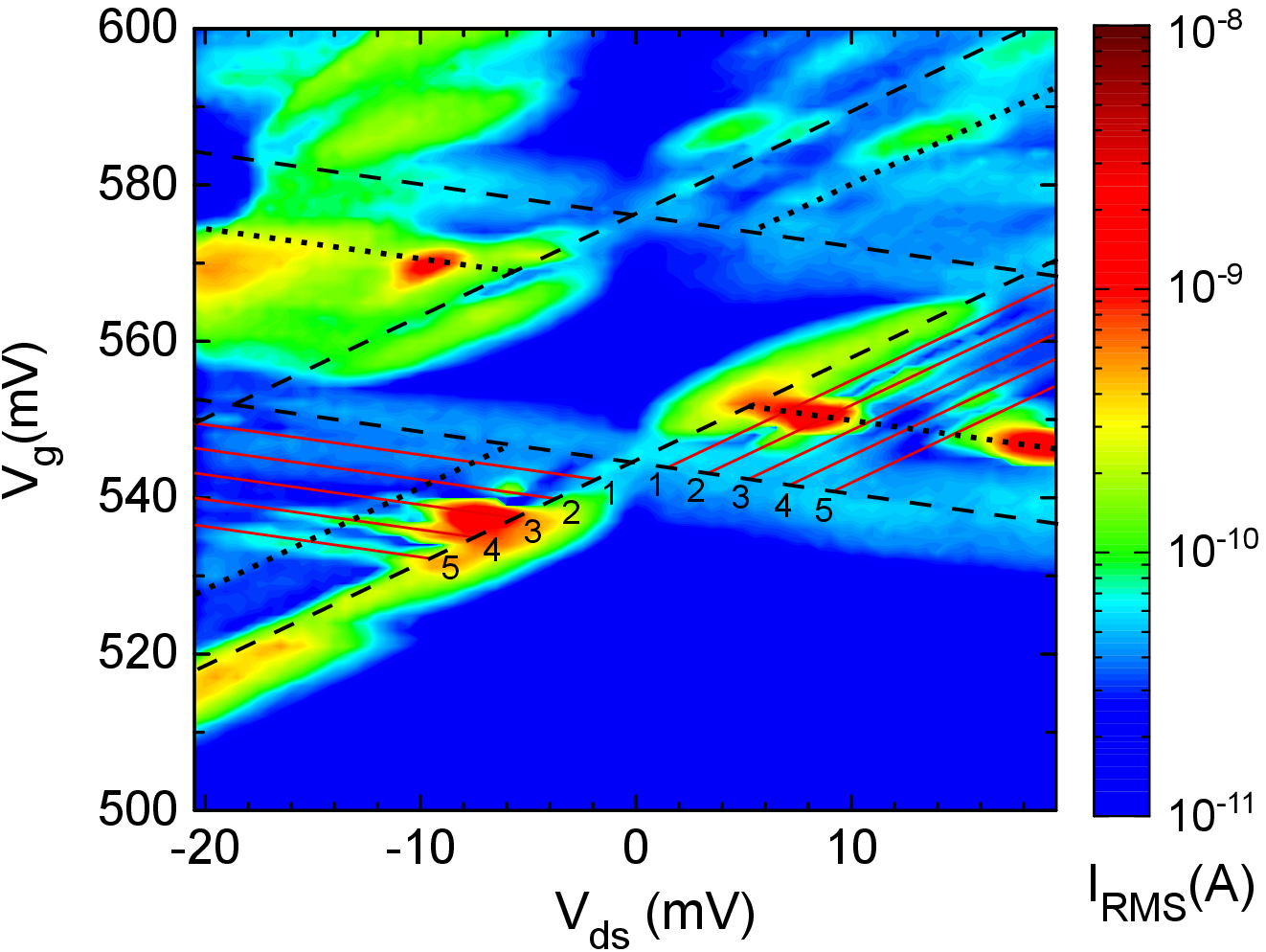}
                            \label{fig:stab-rms}
                            }
                    
                    %\includesvg[width=\linewidth]{2_all.svg}
                    
                    \caption{Stability diagrams of the first Coulomb diamond at $V_b = \SI{0}{\volt}$ and $T = \SI{4.2}{\kelvin}$:
                    \protect\subref{fig:stab-curr} source-drain current,
                    \protect\subref{fig:stab-condg} gate differential transconductance
                    \protect\subref{fig:stab-condd} drain differential conductance
                    \protect\subref{fig:stab-rms} current RMS noise in logarithmic scale,
                    versus gate and drain voltage.
                    Measured points are interpolated along contour lines for better visualization, the actual resolution being $\delta V_{ds} = \SI{0.5}{\milli\volt}$ and $\delta V_g = \SI{1}{\milli\volt}$.
                    The same set of lines is drawn in all four diagrams:
                    black dashed lines highlight the lowest Coulomb diamond, corresponding to the first localized state allowing transport in the device;
                    dotted lines are associated to an excited state about \SI{5}{\milli\electronvolt} above the ground state.
                    The lines marked in red (d) highlight parallel features associated to phonon-mediated transport, their energy separation being approximately around 2.3 meV. The corresponding left and right lines are aligned to the $V_g$ axis. Although they are not all equally visible, the lines highlight the regular spacing.
                }
                
                \label{fig:stab}
            \end{figure*}

        Conductance peaks in \cref{fig:ivtrace} show the resonant tunneling and Coulomb blockade behaviour in the device. The bulk potential $V_b$ is used as a back-gate to tune the electrostatic coupling between the localized states in the channel and the source/drain contacts, which behave as reservoirs with continuous energy spectrum up to their chemical potential \cite{lansbergen2008gate,prati2011adiabatic,prati2013single,prati2016band}.  
        The $V_g$ vs $V_b$ stability diagram at a constant bias of $V_{ds}=1$mV is  reported  in \cref{fig:bulk}. As distinct confinement centers are coupled to the bulk gate differently depending on their position, the resonant tunneling levels result grouped as parallel lines of different slope associated to distinct electrostatic quantum dots and donors centred in the channel.
        The 2D stability diagrams in \cref{fig:stab} of current, differential conductance and differential transconductance versus $V_{ds}$ and $V_g$ give a broader view of the spectroscopy of the states in the device and are complemented by an rms stability diagram.

%%%\subsection{Donor}
        %threshold, ionization energy, charging energy, number of donors?
        The association of the first electron levels of the device to two group V donor atoms is justified by three experimental facts. First, it is the presence of the two pairs of parallel lines (a$_1$, a$_2$ and b$_1$, b$_2$ respectively) in \cref{fig:bulk} with similar charging energy, with no further states at higher energy. It corresponds to the fact that a group V donor can bound only two electrons, forming a neutral ($D^{0}$) and a negatively charged ($D^{-}$) states\cite{prati2013single}.
        %well known fact that no additional electrons can be bound to the donor once negatively charged by the second electron which form a $D^{-}$ state \cite{prati2013single}. 
        Second, the charging energy of both is about 19 meV which is in the range of around 15-20 meV reported in several works\cite{tan2010transport,prati2011adiabatic,pierre10},
        % their relative different voltage range in terms of the gate voltage with respect to those associated to the single electron transistor\cite{leti2011switching},%(Non e` molto chiaro)
        Finally, conduction band transport occurs starting from 40-50 meV above such energy levels, which corresponds to the $\approx 45$meV ionization energy of the ground state referred to the conduction band, which falls around 610 mV at 4.2K as from \cref{fig:ivtrace}. We thus reasonably conclude that, as expected by the fact that the channel is only 100 nm long and the annealing process operated during the fabrication, there are two donors participating to the transport in the channel, both by $D^0$ and $D^-$ states. The values of the charging and ionization energies along with the arsenic implantation for the ohmic contacts definition, suggest the single donors to be As atoms.
        %For what concerns the identification of the species of the donor responsible of the main Coulomb blockade effect, arsenic has been implanted to form the ohmic contacts, so it is likely that this is the species of the single donors.
        At higher energy with respect to the donor states, the family of peaks referred to as \emph{c} have a different slope and energy spacing than the donors, allowing occupation greater than two, which we attribute to a quantum dot electrostatically defined by some interface disorder which is not screened at cryogenic temperature and ultra-diluted electron density.
        Such observations are confirmed by the rigid translation of the stability diagram fingerprint given by the first and the second peaks in \cref{fig:stab} that we observed by changing the bulk voltage.
        %We notice that the two $D^0$ states (a$_1$, b$_1$) appear to exchange roles across the bulk voltage sweep as the current flows across a$_1$ at the lowest $V_b$ and across b$_1$ at the highest. Furthermore they mutually act like anti-lines as previously reported also by some of us \cite{mazzeo12,leti2011switching}.
        An interesting peculiarity of this device consists of its high conductance in units of $e^2/\hbar$ which suggests a strong coupling regime of the donor states with the contacts.

        %The presence of two pairs of peaks at low $V_g$, each with its characteristic slope in the ($V_b, V_g$) stability diagram of \cref{fig:bulk}, suggests the presence of two donor atoms in the channel, each with its $D^0$ and $D^-$ states respectively.
        %The interaction between the levels of the two donors however is different between the two $D^0$ levels than the two $D^-$:   %%%%% COMPLETARE

%%%\subsection{Comparison between the DC current and the RMS noise measurements}\label{sec:stab}
       
        %Negative drain conductance?
        We now turn to the investigation of the electron transport in the device as carried out by first analyzing the Coulomb blockade diamond structures that best fit the data plotted in \cref{fig:stab} according to the constant interaction model \cite{kouwenhoven01}.
        Such measurements were all taken during the same thermal cycle, different than those of \cref{fig:bulk}: a small shift of the resonant tunneling features is observed between different thermal cycles, generally attributed to a different configuration of frozen surrounding charges in distant interface defects, but the relative positions of the peaks and the geometry of the Coulomb diamonds remains the same.
        Figure \ref{fig:stab-rms}) shows the RMS of the drain current as a function of the gate and drain voltages. For each measurement point the acquisition time was of \SI{0.1}{\second} enabling the calculation of the RMS value of the current noise on the full bandwidth of the amplifier starting from \SI{10}{\hertz}.  
        The use of the cryogenic amplifier enhanced the RMS stability diagram measurements, as more resonant tunneling features were observed than in similar diagrams obtained with room-temperature amplification (see Supplementary Material).
        Closely spaced equidistant parallel lines terminating on the lines of the first electron state are detected (red lines in \cref{fig:stab-rms}): such features can not be attributed neither to density-of-state fluctuations in the contacts nor to excited states of the donor because they are symmetric and cross in points of zero bias towards lowest gate voltages respectively.
        Interestingly, their identical periodic spacing for both positive and negative $V_{ds}$ bias voltage points towards tunneling events assisted by phonon emission on the two sides, as from Ref. \citenum{zwanenburg2009spin}. Such transport regime is further investigated in the following by looking at the noise spectrum.

%%%\subsection{Extraction of lever arm factor and capacitances} \label{sec:levarm}
        The width and height of the main Coulomb diamond in \cref{fig:stab}, along with the slopes of its edges and of the red lines in \cref{fig:bulk}, are used to extract information about the donor states and the coupling with the transistor electrodes.
        %In addition, the slope of the red lines in figure \cref{fig:bulk} gives the ratio $C_b/C_g = 0.05$, 
        From the width (i.e. the voltage distance of the corners with respect to the zero bias vertical line) the charging energy is estimated as $E_{ch} = \SI{19\pm 1}{\milli\electronvolt}$.
       
        The lever-arm factor is estimated to be $\alpha = 0.59 \pm 0.05$.
        This value allows to calculate the total capacitance $C$ and the other values as follows:
        
        \begin{center}
        \begin{tabular}{ccccc}
        				%\toprule
        				\hline
        				$C$ & $C_g$ & $C_s$ & $C_d$ & $C_b$ \\
        				%\midrule
        				\hline
        				\SI{13}{\atto\farad} & \SI{7.6}{\atto\farad} & \SI{2.3}{\atto\farad} & \SI{2.5}{\atto\farad} & \SI{0.4}{\atto\farad} \\
        				\hline
        				%\bottomrule
        \end{tabular}
        \end{center}
        Despite the uncertainty in these values due to graphical extrapolation, comparison between them provides insight on the location of the donor atom(s): since $C_s\simeq C_d$, the donor is similarly coupled to the source and drain contacts and therefore located near the center of the channel, as most probable from theory in order to observe some current flow.
        %%spiegazione per C_b?

        The phonon energy spacing $\Delta_{ph}$, again converted from the spacing between the corresponding red lines using the same lever-arm factor is around  $\SI{2.3}{\milli\electronvolt}$.
        Assuming the phonon velocity in silicon of about  $v=\SI{7e5}{\centi\meter/\second}$, %%% origine?
        the estimation of the corresponding wavelength $\lambda $ is of the order of \SI{11}{\nano\meter}.
        In the hypothesis of an open phonon cavity originating these electron transport resonances, as from Ref. \citenum{Escott10}, the length $L$ of such a cavity corresponds to $\lambda/4$;
        the resulting length of 2.6 nm is compatible with the longitudinal Bohr orbit diameter of As atoms in silicon which is 2.36 nm \cite{prati2013single,van2015single}.
        
          \begin{figure}[htp]
                \centering
                \includegraphics[width = \columnwidth]{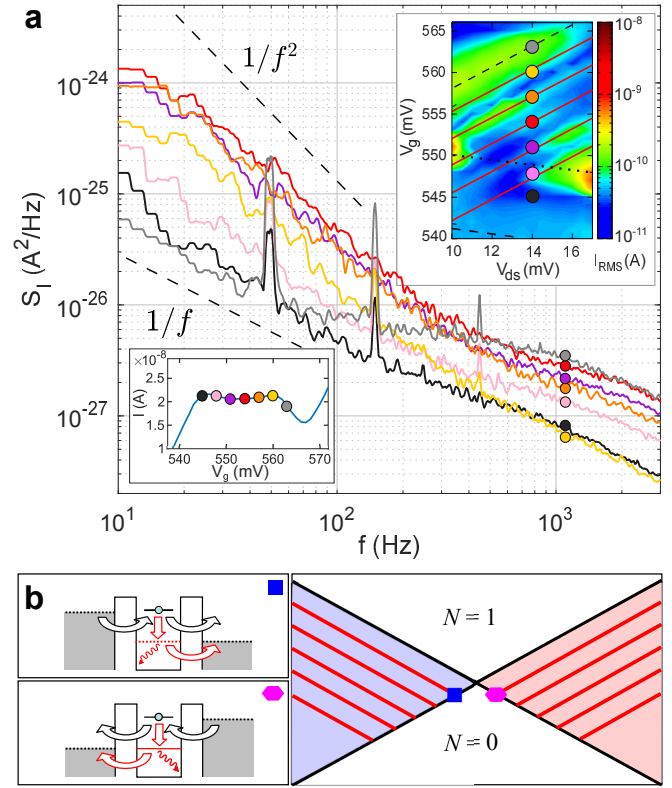}
                \caption{(a) Selected noise spectra at $V_{ds} = \SI{14}{\milli\volt}$, in a region of the stability diagram where evenly spaced phononic lines are observed (see  \protect\cref{fig:stab-rms}). 
                The upper inset zooms the $I_{\mathrm{RMS}}$ data in that portion, measured over the full bandwidth of the acquisition system. The dots indicate the different $V_g$ values at which each respective noise spectrum was measured.
                %The rms integrates a band of about 100 kHz.
                The colored ones correspond to enhanced-current lines at 549, 551, 554, 557 and \SI{560}{\milli\volt} while the black and grey ones, at 543 and \SI{563}{\milli\volt} respectively, where the current flows but the phonons are not involved, are shown for comparison. The dashed lines are guides to the eye of the relevant slopes.
                The peaks observed in the black and grey spectra at 50, 150 and 450 Hz are due to supply current noise.
                The lower inset shows the current at $V_d = 14$ mV indicating the location of the considered working points in a transport region between Coulomb diamonds.
                (b) Sketch of the phonon-mediated transport in correspondence of the first phonon-line. On the left, the mechanism for the negative (blue square) and positive (violet hexagon) current, respectively. On the right, the reference point in the current stability diagram. The red lines indicate where higher RMS noise emerges, not visible by the sole current.}
                \label{fig:spectra}
            \end{figure}
    
%%%%\subsection{Characterization of the phonon-mediated transport regime by the noise spectra}
        Additional information on the phonon-mediated tunneling process has been obtained by measuring the power spectral density of the source-drain current noise. 
        For these measurements the device was tuned to a constant  bias $V_{ds}=14$ mV and to different values of $V_g$ in a range where the evenly-spaced lines associated to phonon-mediated tunneling occurs. 
        \cref{fig:spectra}(a) shows the comparison of the noise spectrum in regions where: current regularly flows through a state at energy far from the transition to the blocked regime, which results as typical $1/f$ noise (black dot), at the edge of the stability diagram where $1/f$ noise is observed at low frequency (gray), and in the regions where phonon emission occurs (yellow, orange, red, purple and pink dots). In this latter case a $1/f^2$ trend is observed according whether the respective lines are visible in the RMS stability diagrams at that bias voltage $V_{DS}$. %\cite{henry1977nonradiative,vitusevich2017noise}. 
        Such low frequency noise can not be associated to the Lorentzian noise of a single defect as the characteristic frequency of the Lorentzian curve fitting the spectra is constant (around 12 Hz) in the whole $V_g$ range of the parallel features, differently than a charge Random Telegraph Signal (RTS). Furthermore, at such low temperature RTS of a single 
        defect arises in energy intervals of few meV only.  
        The spectra of Figure 3a corresponding to phonon lines are fitted by $1/f$ noise with two Lorentzian noise phenomena at 12 Hz and 2800 Hz respectively, both irrespective from the gate voltage. As they are observed only in correspondence of visible phonon emission lines, we attribute them to the decay of high frequency phonons to low frequency acoustic phonons during the cooling process of the crystal \cite{galkina1987down,klitsner1987phonon,msall1993observation,xie2013phonon}.
        %There, at low frequency, the noise power spectral density exhibits additionally a  $1/f^2$ dependence, which is compatible with the tail of another Lorentzian spectrum associated whose origin can be attributed to the relaxation of the crystal after absorbing the phonons generated by the quantum transport. 
        The \cref{fig:spectra}(b) shows the microscopic mechanism of phonon emission for the negative and positive current respectively. To summarize, the single phonon-mediated regime of the quantum transport occurs in such single electron filling of the single donor channel regime.

%%%section{Conclusion}
To conclude, we observe single phonon-mediated quantum transport by tracking the RMS noise in a nano-transistor operated in the single atom and single electron regime. The use of a cryogenic amplifier makes such measurement effective, by reducing the instrumental noise and the electromagnetic noise affecting the measurements when the amplification system is far from the device under test. The characterization of a commercial single electron transistor revealed the presence of a dopant in the channel, likely originated by unintentional diffusion from the source/drain contacts. The RMS measurements indicate the presence of few- down to single-phonon mediated tunneling, for which the low frequency spectrum results a Lorentzian typical of generation-recombination processes.   

\medskip
\noindent
Data are available upon reasonable request.
\bibliography{bibliography}

\begin{thebibliography}{29}%
\makeatletter
\providecommand \@ifxundefined [1]{%
 \@ifx{#1\undefined}
}%
\providecommand \@ifnum [1]{%
 \ifnum #1\expandafter \@firstoftwo
 \else \expandafter \@secondoftwo
 \fi
}%
\providecommand \@ifx [1]{%
 \ifx #1\expandafter \@firstoftwo
 \else \expandafter \@secondoftwo
 \fi
}%
\providecommand \natexlab [1]{#1}%
\providecommand \enquote  [1]{``#1''}%
\providecommand \bibnamefont  [1]{#1}%
\providecommand \bibfnamefont [1]{#1}%
\providecommand \citenamefont [1]{#1}%
\providecommand \href@noop [0]{\@secondoftwo}%
\providecommand \href [0]{\begingroup \@sanitize@url \@href}%
\providecommand \@href[1]{\@@startlink{#1}\@@href}%
\providecommand \@@href[1]{\endgroup#1\@@endlink}%
\providecommand \@sanitize@url [0]{\catcode `\\12\catcode `\$12\catcode
  `\&12\catcode `\#12\catcode `\^12\catcode `\_12\catcode `\%12\relax}%
\providecommand \@@startlink[1]{}%
\providecommand \@@endlink[0]{}%
\providecommand \url  [0]{\begingroup\@sanitize@url \@url }%
\providecommand \@url [1]{\endgroup\@href {#1}{\urlprefix }}%
\providecommand \urlprefix  [0]{URL }%
\providecommand \Eprint [0]{\href }%
\providecommand \doibase [0]{http://dx.doi.org/}%
\providecommand \selectlanguage [0]{\@gobble}%
\providecommand \bibinfo  [0]{\@secondoftwo}%
\providecommand \bibfield  [0]{\@secondoftwo}%
\providecommand \translation [1]{[#1]}%
\providecommand \BibitemOpen [0]{}%
\providecommand \bibitemStop [0]{}%
\providecommand \bibitemNoStop [0]{.\EOS\space}%
\providecommand \EOS [0]{\spacefactor3000\relax}%
\providecommand \BibitemShut  [1]{\csname bibitem#1\endcsname}%
\let\auto@bib@innerbib\@empty
%</preamble>
\bibitem [{\citenamefont {Fujisawa}\ \emph {et~al.}(1998)\citenamefont
  {Fujisawa}, \citenamefont {Oosterkamp}, \citenamefont {Van~der Wiel},
  \citenamefont {Broer}, \citenamefont {Aguado}, \citenamefont {Tarucha},\ and\
  \citenamefont {Kouwenhoven}}]{fujisawa1998spontaneous}%
  \BibitemOpen
  \bibfield  {author} {\bibinfo {author} {\bibfnamefont {T.}~\bibnamefont
  {Fujisawa}}, \bibinfo {author} {\bibfnamefont {T.~H.}\ \bibnamefont
  {Oosterkamp}}, \bibinfo {author} {\bibfnamefont {W.~G.}\ \bibnamefont
  {Van~der Wiel}}, \bibinfo {author} {\bibfnamefont {B.~W.}\ \bibnamefont
  {Broer}}, \bibinfo {author} {\bibfnamefont {R.}~\bibnamefont {Aguado}},
  \bibinfo {author} {\bibfnamefont {S.}~\bibnamefont {Tarucha}}, \ and\
  \bibinfo {author} {\bibfnamefont {L.~P.}\ \bibnamefont {Kouwenhoven}},\
  }\href@noop {} {\bibfield  {journal} {\bibinfo  {journal} {Science}\ }\textbf
  {\bibinfo {volume} {282}},\ \bibinfo {pages} {932} (\bibinfo {year}
  {1998})}\BibitemShut {NoStop}%
\bibitem [{\citenamefont {Zwanenburg}\ \emph {et~al.}(2009)\citenamefont
  {Zwanenburg}, \citenamefont {van Rijmenam}, \citenamefont {Fang},
  \citenamefont {Lieber},\ and\ \citenamefont
  {Kouwenhoven}}]{zwanenburg2009spin}%
  \BibitemOpen
  \bibfield  {author} {\bibinfo {author} {\bibfnamefont {F.~A.}\ \bibnamefont
  {Zwanenburg}}, \bibinfo {author} {\bibfnamefont {C.~E.}\ \bibnamefont {van
  Rijmenam}}, \bibinfo {author} {\bibfnamefont {Y.}~\bibnamefont {Fang}},
  \bibinfo {author} {\bibfnamefont {C.~M.}\ \bibnamefont {Lieber}}, \ and\
  \bibinfo {author} {\bibfnamefont {L.~P.}\ \bibnamefont {Kouwenhoven}},\
  }\href@noop {} {\bibfield  {journal} {\bibinfo  {journal} {Nano letters}\
  }\textbf {\bibinfo {volume} {9}},\ \bibinfo {pages} {1071} (\bibinfo {year}
  {2009})}\BibitemShut {NoStop}%
\bibitem [{\citenamefont {Escott}, \citenamefont {Zwanenburg},\ and\
  \citenamefont {Morello}(2010)}]{Escott10}%
  \BibitemOpen
  \bibfield  {author} {\bibinfo {author} {\bibfnamefont {C.~C.}\ \bibnamefont
  {Escott}}, \bibinfo {author} {\bibfnamefont {F.~A.}\ \bibnamefont
  {Zwanenburg}}, \ and\ \bibinfo {author} {\bibfnamefont {A.}~\bibnamefont
  {Morello}},\ }\href {\doibase 10.1088/0957-4484/21/27/274018} {\bibfield
  {journal} {\bibinfo  {journal} {Nanotechnology}\ }\textbf {\bibinfo {volume}
  {21}},\ \bibinfo {pages} {274018} (\bibinfo {year} {2010})}\BibitemShut
  {NoStop}%
\bibitem [{\citenamefont {Granger}\ \emph {et~al.}(2012)\citenamefont
  {Granger}, \citenamefont {Taubert}, \citenamefont {Young}, \citenamefont
  {Gaudreau}, \citenamefont {Kam}, \citenamefont {Studenikin}, \citenamefont
  {Zawadzki}, \citenamefont {Harbusch}, \citenamefont {Schuh}, \citenamefont
  {Wegscheider} \emph {et~al.}}]{granger2012quantum}%
  \BibitemOpen
  \bibfield  {author} {\bibinfo {author} {\bibfnamefont {G.}~\bibnamefont
  {Granger}}, \bibinfo {author} {\bibfnamefont {D.}~\bibnamefont {Taubert}},
  \bibinfo {author} {\bibfnamefont {C.}~\bibnamefont {Young}}, \bibinfo
  {author} {\bibfnamefont {L.}~\bibnamefont {Gaudreau}}, \bibinfo {author}
  {\bibfnamefont {A.}~\bibnamefont {Kam}}, \bibinfo {author} {\bibfnamefont
  {S.}~\bibnamefont {Studenikin}}, \bibinfo {author} {\bibfnamefont
  {P.}~\bibnamefont {Zawadzki}}, \bibinfo {author} {\bibfnamefont
  {D.}~\bibnamefont {Harbusch}}, \bibinfo {author} {\bibfnamefont
  {D.}~\bibnamefont {Schuh}}, \bibinfo {author} {\bibfnamefont
  {W.}~\bibnamefont {Wegscheider}},  \emph {et~al.},\ }\href@noop {} {\bibfield
   {journal} {\bibinfo  {journal} {Nature Physics}\ }\textbf {\bibinfo {volume}
  {8}},\ \bibinfo {pages} {522} (\bibinfo {year} {2012})}\BibitemShut {NoStop}%
\bibitem [{\citenamefont {Braakman}\ \emph {et~al.}(2013)\citenamefont
  {Braakman}, \citenamefont {Barthelemy}, \citenamefont {Reichl}, \citenamefont
  {Wegscheider},\ and\ \citenamefont {Vandersypen}}]{braakman2013photon}%
  \BibitemOpen
  \bibfield  {author} {\bibinfo {author} {\bibfnamefont {F.~R.}\ \bibnamefont
  {Braakman}}, \bibinfo {author} {\bibfnamefont {P.}~\bibnamefont
  {Barthelemy}}, \bibinfo {author} {\bibfnamefont {C.}~\bibnamefont {Reichl}},
  \bibinfo {author} {\bibfnamefont {W.}~\bibnamefont {Wegscheider}}, \ and\
  \bibinfo {author} {\bibfnamefont {L.~M.}\ \bibnamefont {Vandersypen}},\
  }\href@noop {} {\bibfield  {journal} {\bibinfo  {journal} {Applied Physics
  Letters}\ }\textbf {\bibinfo {volume} {102}},\ \bibinfo {pages} {112110}
  (\bibinfo {year} {2013})}\BibitemShut {NoStop}%
\bibitem [{\citenamefont {Braig}\ and\ \citenamefont
  {Flensberg}(2003)}]{braig2003vibrational}%
  \BibitemOpen
  \bibfield  {author} {\bibinfo {author} {\bibfnamefont {S.}~\bibnamefont
  {Braig}}\ and\ \bibinfo {author} {\bibfnamefont {K.}~\bibnamefont
  {Flensberg}},\ }\href@noop {} {\bibfield  {journal} {\bibinfo  {journal}
  {Physical Review B}\ }\textbf {\bibinfo {volume} {68}},\ \bibinfo {pages}
  {205324} (\bibinfo {year} {2003})}\BibitemShut {NoStop}%
\bibitem [{\citenamefont {Crippa}\ \emph {et~al.}(2019)\citenamefont {Crippa},
  \citenamefont {Ezzouch}, \citenamefont {Apr{\'a}}, \citenamefont {Amisse},
  \citenamefont {Lavi{\'e}ville}, \citenamefont {Hutin}, \citenamefont
  {Bertrand}, \citenamefont {Vinet}, \citenamefont {Urdampilleta},
  \citenamefont {Meunier} \emph {et~al.}}]{crippa2019gate}%
  \BibitemOpen
  \bibfield  {author} {\bibinfo {author} {\bibfnamefont {A.}~\bibnamefont
  {Crippa}}, \bibinfo {author} {\bibfnamefont {R.}~\bibnamefont {Ezzouch}},
  \bibinfo {author} {\bibfnamefont {A.}~\bibnamefont {Apr{\'a}}}, \bibinfo
  {author} {\bibfnamefont {A.}~\bibnamefont {Amisse}}, \bibinfo {author}
  {\bibfnamefont {R.}~\bibnamefont {Lavi{\'e}ville}}, \bibinfo {author}
  {\bibfnamefont {L.}~\bibnamefont {Hutin}}, \bibinfo {author} {\bibfnamefont
  {B.}~\bibnamefont {Bertrand}}, \bibinfo {author} {\bibfnamefont
  {M.}~\bibnamefont {Vinet}}, \bibinfo {author} {\bibfnamefont
  {M.}~\bibnamefont {Urdampilleta}}, \bibinfo {author} {\bibfnamefont
  {T.}~\bibnamefont {Meunier}},  \emph {et~al.},\ }\href@noop {} {\bibfield
  {journal} {\bibinfo  {journal} {Nature communications}\ }\textbf {\bibinfo
  {volume} {10}},\ \bibinfo {pages} {1} (\bibinfo {year} {2019})}\BibitemShut
  {NoStop}%
\bibitem [{\citenamefont {Turchetti}\ \emph {et~al.}(2015)\citenamefont
  {Turchetti}, \citenamefont {Homulle}, \citenamefont {Sebastiano},
  \citenamefont {Ferrari}, \citenamefont {Charbon},\ and\ \citenamefont
  {Prati}}]{turchetti15}%
  \BibitemOpen
  \bibfield  {author} {\bibinfo {author} {\bibfnamefont {M.}~\bibnamefont
  {Turchetti}}, \bibinfo {author} {\bibfnamefont {H.}~\bibnamefont {Homulle}},
  \bibinfo {author} {\bibfnamefont {F.}~\bibnamefont {Sebastiano}}, \bibinfo
  {author} {\bibfnamefont {G.}~\bibnamefont {Ferrari}}, \bibinfo {author}
  {\bibfnamefont {E.}~\bibnamefont {Charbon}}, \ and\ \bibinfo {author}
  {\bibfnamefont {E.}~\bibnamefont {Prati}},\ }\href@noop {} {\bibfield
  {journal} {\bibinfo  {journal} {Applied Physics Express}\ }\textbf {\bibinfo
  {volume} {9}},\ \bibinfo {pages} {014001} (\bibinfo {year}
  {2015})}\BibitemShut {NoStop}%
\bibitem [{\citenamefont {Leti}\ \emph {et~al.}(2011)\citenamefont {Leti},
  \citenamefont {Prati}, \citenamefont {Belli}, \citenamefont {Petretto},
  \citenamefont {Fanciulli}, \citenamefont {Vinet}, \citenamefont {Wacquez},\
  and\ \citenamefont {Sanquer}}]{leti2011switching}%
  \BibitemOpen
  \bibfield  {author} {\bibinfo {author} {\bibfnamefont {G.}~\bibnamefont
  {Leti}}, \bibinfo {author} {\bibfnamefont {E.}~\bibnamefont {Prati}},
  \bibinfo {author} {\bibfnamefont {M.}~\bibnamefont {Belli}}, \bibinfo
  {author} {\bibfnamefont {G.}~\bibnamefont {Petretto}}, \bibinfo {author}
  {\bibfnamefont {M.}~\bibnamefont {Fanciulli}}, \bibinfo {author}
  {\bibfnamefont {M.}~\bibnamefont {Vinet}}, \bibinfo {author} {\bibfnamefont
  {R.}~\bibnamefont {Wacquez}}, \ and\ \bibinfo {author} {\bibfnamefont
  {M.}~\bibnamefont {Sanquer}},\ }\href@noop {} {\bibfield  {journal} {\bibinfo
   {journal} {Applied Physics Letters}\ }\textbf {\bibinfo {volume} {99}},\
  \bibinfo {pages} {242102} (\bibinfo {year} {2011})}\BibitemShut {NoStop}%
\bibitem [{\citenamefont {Crippa}\ \emph {et~al.}(2015)\citenamefont {Crippa},
  \citenamefont {Tagliaferri}, \citenamefont {Rotta}, \citenamefont {{De
  Michielis}}, \citenamefont {Mazzeo}, \citenamefont {Fanciulli}, \citenamefont
  {Wacquez}, \citenamefont {Vinet},\ and\ \citenamefont
  {Prati}}]{crippa2015valley}%
  \BibitemOpen
  \bibfield  {author} {\bibinfo {author} {\bibfnamefont {A.}~\bibnamefont
  {Crippa}}, \bibinfo {author} {\bibfnamefont {M.~L.}\ \bibnamefont
  {Tagliaferri}}, \bibinfo {author} {\bibfnamefont {D.}~\bibnamefont {Rotta}},
  \bibinfo {author} {\bibfnamefont {M.}~\bibnamefont {{De Michielis}}},
  \bibinfo {author} {\bibfnamefont {G.}~\bibnamefont {Mazzeo}}, \bibinfo
  {author} {\bibfnamefont {M.}~\bibnamefont {Fanciulli}}, \bibinfo {author}
  {\bibfnamefont {R.}~\bibnamefont {Wacquez}}, \bibinfo {author} {\bibfnamefont
  {M.}~\bibnamefont {Vinet}}, \ and\ \bibinfo {author} {\bibfnamefont
  {E.}~\bibnamefont {Prati}},\ }\href {\doibase 10.1103/PhysRevB.92.035424}
  {\bibfield  {journal} {\bibinfo  {journal} {Physical Review B}\ }\textbf
  {\bibinfo {volume} {92}},\ \bibinfo {pages} {35424} (\bibinfo {year}
  {2015})}\BibitemShut {NoStop}%
\bibitem [{\citenamefont {Guagliardo}\ and\ \citenamefont
  {Ferrari}(2013)}]{Guagliardo2013}%
  \BibitemOpen
  \bibfield  {author} {\bibinfo {author} {\bibfnamefont {F.}~\bibnamefont
  {Guagliardo}}\ and\ \bibinfo {author} {\bibfnamefont {G.}~\bibnamefont
  {Ferrari}},\ }in\ \href {\doibase 10.4032/9789814316699} {\emph {\bibinfo
  {booktitle} {Low-noise current measurements on quantum devices operating at
  cryogenic temperature}}},\ \bibinfo {editor} {edited by\ \bibinfo {editor}
  {\bibfnamefont {E.}~\bibnamefont {Prati}}\ and\ \bibinfo {editor}
  {\bibfnamefont {T.}~\bibnamefont {Shinada}}}\ (\bibinfo  {publisher} {Pan
  Stanford Publishing},\ \bibinfo {year} {2013})\ pp.\ \bibinfo {pages}
  {187--210}\BibitemShut {NoStop}%
\bibitem [{\citenamefont {Tagliaferri}\ \emph
  {et~al.}(2016{\natexlab{a}})\citenamefont {Tagliaferri}, \citenamefont
  {Crippa}, \citenamefont {Cocco}, \citenamefont {De~Michielis}, \citenamefont
  {Fanciulli}, \citenamefont {Ferrari},\ and\ \citenamefont
  {Prati}}]{tagliaferri16}%
  \BibitemOpen
  \bibfield  {author} {\bibinfo {author} {\bibfnamefont {M.~L.~V.}\
  \bibnamefont {Tagliaferri}}, \bibinfo {author} {\bibfnamefont
  {A.}~\bibnamefont {Crippa}}, \bibinfo {author} {\bibfnamefont
  {S.}~\bibnamefont {Cocco}}, \bibinfo {author} {\bibfnamefont
  {M.}~\bibnamefont {De~Michielis}}, \bibinfo {author} {\bibfnamefont
  {M.}~\bibnamefont {Fanciulli}}, \bibinfo {author} {\bibfnamefont
  {G.}~\bibnamefont {Ferrari}}, \ and\ \bibinfo {author} {\bibfnamefont
  {E.}~\bibnamefont {Prati}},\ }\href@noop {} {\bibfield  {journal} {\bibinfo
  {journal} {IEEE Transactions on Instrumentation and Measurement}\ }\textbf
  {\bibinfo {volume} {65}},\ \bibinfo {pages} {1827} (\bibinfo {year}
  {2016}{\natexlab{a}})}\BibitemShut {NoStop}%
\bibitem [{\citenamefont {Prati}\ and\ \citenamefont
  {Morello}(2013)}]{prati2013quantum}%
  \BibitemOpen
  \bibfield  {author} {\bibinfo {author} {\bibfnamefont {E.}~\bibnamefont
  {Prati}}\ and\ \bibinfo {author} {\bibfnamefont {A.}~\bibnamefont
  {Morello}},\ }\href@noop {} {\bibfield  {journal} {\bibinfo  {journal}
  {Single-Atom Nanoelectronics}\ ,\ \bibinfo {pages} {5}} (\bibinfo {year}
  {2013})}\BibitemShut {NoStop}%
\bibitem [{\citenamefont {Vitusevich}\ and\ \citenamefont
  {Zadorozhnyi}(2017)}]{vitusevich2017noise}%
  \BibitemOpen
  \bibfield  {author} {\bibinfo {author} {\bibfnamefont {S.}~\bibnamefont
  {Vitusevich}}\ and\ \bibinfo {author} {\bibfnamefont {I.}~\bibnamefont
  {Zadorozhnyi}},\ }\href@noop {} {\bibfield  {journal} {\bibinfo  {journal}
  {Semiconductor Science and Technology}\ }\textbf {\bibinfo {volume} {32}},\
  \bibinfo {pages} {043002} (\bibinfo {year} {2017})}\BibitemShut {NoStop}%
\bibitem [{\citenamefont {Mazzeo}\ \emph {et~al.}(2012)\citenamefont {Mazzeo},
  \citenamefont {Prati}, \citenamefont {Belli}, \citenamefont {Leti},
  \citenamefont {Cocco}, \citenamefont {Fanciulli}, \citenamefont
  {Guagliardo},\ and\ \citenamefont {Ferrari}}]{mazzeo12}%
  \BibitemOpen
  \bibfield  {author} {\bibinfo {author} {\bibfnamefont {G.}~\bibnamefont
  {Mazzeo}}, \bibinfo {author} {\bibfnamefont {E.}~\bibnamefont {Prati}},
  \bibinfo {author} {\bibfnamefont {M.}~\bibnamefont {Belli}}, \bibinfo
  {author} {\bibfnamefont {G.}~\bibnamefont {Leti}}, \bibinfo {author}
  {\bibfnamefont {S.}~\bibnamefont {Cocco}}, \bibinfo {author} {\bibfnamefont
  {M.}~\bibnamefont {Fanciulli}}, \bibinfo {author} {\bibfnamefont
  {F.}~\bibnamefont {Guagliardo}}, \ and\ \bibinfo {author} {\bibfnamefont
  {G.}~\bibnamefont {Ferrari}},\ }\href@noop {} {\bibfield  {journal} {\bibinfo
   {journal} {Applied Physics Letters}\ }\textbf {\bibinfo {volume} {100}},\
  \bibinfo {pages} {213107} (\bibinfo {year} {2012})}\BibitemShut {NoStop}%
\bibitem [{\citenamefont {Tagliaferri}\ \emph
  {et~al.}(2016{\natexlab{b}})\citenamefont {Tagliaferri}, \citenamefont
  {Crippa}, \citenamefont {De~Michielis}, \citenamefont {Mazzeo}, \citenamefont
  {Fanciulli},\ and\ \citenamefont {Prati}}]{tagliaferri2016compact}%
  \BibitemOpen
  \bibfield  {author} {\bibinfo {author} {\bibfnamefont {M.~L.~V.}\
  \bibnamefont {Tagliaferri}}, \bibinfo {author} {\bibfnamefont
  {A.}~\bibnamefont {Crippa}}, \bibinfo {author} {\bibfnamefont
  {M.}~\bibnamefont {De~Michielis}}, \bibinfo {author} {\bibfnamefont
  {G.}~\bibnamefont {Mazzeo}}, \bibinfo {author} {\bibfnamefont
  {M.}~\bibnamefont {Fanciulli}}, \ and\ \bibinfo {author} {\bibfnamefont
  {E.}~\bibnamefont {Prati}},\ }\href@noop {} {\bibfield  {journal} {\bibinfo
  {journal} {Physics Letters A}\ }\textbf {\bibinfo {volume} {380}},\ \bibinfo
  {pages} {1205} (\bibinfo {year} {2016}{\natexlab{b}})}\BibitemShut {NoStop}%
\bibitem [{\citenamefont {Sampietro}, \citenamefont {Fasoli},\ and\
  \citenamefont {Ferrari}(1999)}]{sampietro1999}%
  \BibitemOpen
  \bibfield  {author} {\bibinfo {author} {\bibfnamefont {M.}~\bibnamefont
  {Sampietro}}, \bibinfo {author} {\bibfnamefont {L.}~\bibnamefont {Fasoli}}, \
  and\ \bibinfo {author} {\bibfnamefont {G.}~\bibnamefont {Ferrari}},\
  }\href@noop {} {\bibfield  {journal} {\bibinfo  {journal} {Review of
  scientific instruments}\ }\textbf {\bibinfo {volume} {70}},\ \bibinfo {pages}
  {2520} (\bibinfo {year} {1999})}\BibitemShut {NoStop}%
\bibitem [{\citenamefont {Lansbergen}\ \emph {et~al.}(2008)\citenamefont
  {Lansbergen}, \citenamefont {Rahman}, \citenamefont {Wellard}, \citenamefont
  {Woo}, \citenamefont {Caro}, \citenamefont {Collaert}, \citenamefont
  {Biesemans}, \citenamefont {Klimeck}, \citenamefont {Hollenberg},\ and\
  \citenamefont {Rogge}}]{lansbergen2008gate}%
  \BibitemOpen
  \bibfield  {author} {\bibinfo {author} {\bibfnamefont {G.}~\bibnamefont
  {Lansbergen}}, \bibinfo {author} {\bibfnamefont {R.}~\bibnamefont {Rahman}},
  \bibinfo {author} {\bibfnamefont {C.}~\bibnamefont {Wellard}}, \bibinfo
  {author} {\bibfnamefont {I.}~\bibnamefont {Woo}}, \bibinfo {author}
  {\bibfnamefont {J.}~\bibnamefont {Caro}}, \bibinfo {author} {\bibfnamefont
  {N.}~\bibnamefont {Collaert}}, \bibinfo {author} {\bibfnamefont
  {S.}~\bibnamefont {Biesemans}}, \bibinfo {author} {\bibfnamefont
  {G.}~\bibnamefont {Klimeck}}, \bibinfo {author} {\bibfnamefont
  {L.}~\bibnamefont {Hollenberg}}, \ and\ \bibinfo {author} {\bibfnamefont
  {S.}~\bibnamefont {Rogge}},\ }\href@noop {} {\bibfield  {journal} {\bibinfo
  {journal} {Nature Physics}\ }\textbf {\bibinfo {volume} {4}},\ \bibinfo
  {pages} {656} (\bibinfo {year} {2008})}\BibitemShut {NoStop}%
\bibitem [{\citenamefont {Prati}\ \emph {et~al.}(2011)\citenamefont {Prati},
  \citenamefont {Belli}, \citenamefont {Cocco}, \citenamefont {Petretto},\ and\
  \citenamefont {Fanciulli}}]{prati2011adiabatic}%
  \BibitemOpen
  \bibfield  {author} {\bibinfo {author} {\bibfnamefont {E.}~\bibnamefont
  {Prati}}, \bibinfo {author} {\bibfnamefont {M.}~\bibnamefont {Belli}},
  \bibinfo {author} {\bibfnamefont {S.}~\bibnamefont {Cocco}}, \bibinfo
  {author} {\bibfnamefont {G.}~\bibnamefont {Petretto}}, \ and\ \bibinfo
  {author} {\bibfnamefont {M.}~\bibnamefont {Fanciulli}},\ }\href@noop {}
  {\bibfield  {journal} {\bibinfo  {journal} {Applied physics letters}\
  }\textbf {\bibinfo {volume} {98}},\ \bibinfo {pages} {053109} (\bibinfo
  {year} {2011})}\BibitemShut {NoStop}%
\bibitem [{\citenamefont {Prati}\ and\ \citenamefont
  {Shinada}(2013)}]{prati2013single}%
  \BibitemOpen
  \bibfield  {author} {\bibinfo {author} {\bibfnamefont {E.}~\bibnamefont
  {Prati}}\ and\ \bibinfo {author} {\bibfnamefont {T.}~\bibnamefont
  {Shinada}},\ }\href@noop {} {\emph {\bibinfo {title} {Single-atom
  nanoelectronics}}}\ (\bibinfo  {publisher} {Jenny Stanford Publishing},\
  \bibinfo {year} {2013})\ p.~\bibinfo {pages} {16}\BibitemShut {NoStop}%
\bibitem [{\citenamefont {Prati}\ \emph {et~al.}(2016)\citenamefont {Prati},
  \citenamefont {Kumagai}, \citenamefont {Hori},\ and\ \citenamefont
  {Shinada}}]{prati2016band}%
  \BibitemOpen
  \bibfield  {author} {\bibinfo {author} {\bibfnamefont {E.}~\bibnamefont
  {Prati}}, \bibinfo {author} {\bibfnamefont {K.}~\bibnamefont {Kumagai}},
  \bibinfo {author} {\bibfnamefont {M.}~\bibnamefont {Hori}}, \ and\ \bibinfo
  {author} {\bibfnamefont {T.}~\bibnamefont {Shinada}},\ }\href@noop {}
  {\bibfield  {journal} {\bibinfo  {journal} {Scientific reports}\ }\textbf
  {\bibinfo {volume} {6}},\ \bibinfo {pages} {19704} (\bibinfo {year}
  {2016})}\BibitemShut {NoStop}%
\bibitem [{\citenamefont {Tan}\ \emph {et~al.}(2010)\citenamefont {Tan},
  \citenamefont {Chan}, \citenamefont {Mottonen}, \citenamefont {Morello},
  \citenamefont {Yang}, \citenamefont {Donkelaar}, \citenamefont {Alves},
  \citenamefont {Pirkkalainen}, \citenamefont {Jamieson}, \citenamefont {Clark}
  \emph {et~al.}}]{tan2010transport}%
  \BibitemOpen
  \bibfield  {author} {\bibinfo {author} {\bibfnamefont {K.~Y.}\ \bibnamefont
  {Tan}}, \bibinfo {author} {\bibfnamefont {K.~W.}\ \bibnamefont {Chan}},
  \bibinfo {author} {\bibfnamefont {M.}~\bibnamefont {Mottonen}}, \bibinfo
  {author} {\bibfnamefont {A.}~\bibnamefont {Morello}}, \bibinfo {author}
  {\bibfnamefont {C.}~\bibnamefont {Yang}}, \bibinfo {author} {\bibfnamefont
  {J.~v.}\ \bibnamefont {Donkelaar}}, \bibinfo {author} {\bibfnamefont
  {A.}~\bibnamefont {Alves}}, \bibinfo {author} {\bibfnamefont {J.-M.}\
  \bibnamefont {Pirkkalainen}}, \bibinfo {author} {\bibfnamefont {D.~N.}\
  \bibnamefont {Jamieson}}, \bibinfo {author} {\bibfnamefont {R.~G.}\
  \bibnamefont {Clark}},  \emph {et~al.},\ }\href@noop {} {\bibfield  {journal}
  {\bibinfo  {journal} {Nano letters}\ }\textbf {\bibinfo {volume} {10}},\
  \bibinfo {pages} {11} (\bibinfo {year} {2010})}\BibitemShut {NoStop}%
\bibitem [{\citenamefont {Pierre}\ \emph {et~al.}(2010)\citenamefont {Pierre},
  \citenamefont {Wacquez}, \citenamefont {Jehl}, \citenamefont {Sanquer},
  \citenamefont {Vinet},\ and\ \citenamefont {Cueto}}]{pierre10}%
  \BibitemOpen
  \bibfield  {author} {\bibinfo {author} {\bibfnamefont {M.}~\bibnamefont
  {Pierre}}, \bibinfo {author} {\bibfnamefont {R.}~\bibnamefont {Wacquez}},
  \bibinfo {author} {\bibfnamefont {X.}~\bibnamefont {Jehl}}, \bibinfo {author}
  {\bibfnamefont {M.}~\bibnamefont {Sanquer}}, \bibinfo {author} {\bibfnamefont
  {M.}~\bibnamefont {Vinet}}, \ and\ \bibinfo {author} {\bibfnamefont
  {O.}~\bibnamefont {Cueto}},\ }\href@noop {} {\bibfield  {journal} {\bibinfo
  {journal} {Nature nanotechnology}\ }\textbf {\bibinfo {volume} {5}},\
  \bibinfo {pages} {133} (\bibinfo {year} {2010})}\BibitemShut {NoStop}%
\bibitem [{\citenamefont {Kouwenhoven}, \citenamefont {Austing},\ and\
  \citenamefont {Tarucha}(2001)}]{kouwenhoven01}%
  \BibitemOpen
  \bibfield  {author} {\bibinfo {author} {\bibfnamefont {L.~P.}\ \bibnamefont
  {Kouwenhoven}}, \bibinfo {author} {\bibfnamefont {D.}~\bibnamefont
  {Austing}}, \ and\ \bibinfo {author} {\bibfnamefont {S.}~\bibnamefont
  {Tarucha}},\ }\href@noop {} {\bibfield  {journal} {\bibinfo  {journal}
  {Reports on Progress in Physics}\ }\textbf {\bibinfo {volume} {64}},\
  \bibinfo {pages} {701} (\bibinfo {year} {2001})}\BibitemShut {NoStop}%
\bibitem [{\citenamefont {Van~Donkelaar}\ \emph {et~al.}(2015)\citenamefont
  {Van~Donkelaar}, \citenamefont {Yang}, \citenamefont {Alves}, \citenamefont
  {McCallum}, \citenamefont {Hougaard}, \citenamefont {Johnson}, \citenamefont
  {Hudson}, \citenamefont {Dzurak}, \citenamefont {Morello}, \citenamefont
  {Spemann} \emph {et~al.}}]{van2015single}%
  \BibitemOpen
  \bibfield  {author} {\bibinfo {author} {\bibfnamefont {J.}~\bibnamefont
  {Van~Donkelaar}}, \bibinfo {author} {\bibfnamefont {C.}~\bibnamefont {Yang}},
  \bibinfo {author} {\bibfnamefont {A.}~\bibnamefont {Alves}}, \bibinfo
  {author} {\bibfnamefont {J.}~\bibnamefont {McCallum}}, \bibinfo {author}
  {\bibfnamefont {C.}~\bibnamefont {Hougaard}}, \bibinfo {author}
  {\bibfnamefont {B.}~\bibnamefont {Johnson}}, \bibinfo {author} {\bibfnamefont
  {F.}~\bibnamefont {Hudson}}, \bibinfo {author} {\bibfnamefont
  {A.}~\bibnamefont {Dzurak}}, \bibinfo {author} {\bibfnamefont
  {A.}~\bibnamefont {Morello}}, \bibinfo {author} {\bibfnamefont
  {D.}~\bibnamefont {Spemann}},  \emph {et~al.},\ }\href@noop {} {\bibfield
  {journal} {\bibinfo  {journal} {Journal of Physics: Condensed Matter}\
  }\textbf {\bibinfo {volume} {27}},\ \bibinfo {pages} {154204} (\bibinfo
  {year} {2015})}\BibitemShut {NoStop}%
\bibitem [{\citenamefont {Galkina}\ \emph {et~al.}(1987)\citenamefont
  {Galkina}, \citenamefont {Blinov}, \citenamefont {Bonch-Osmolovskii},
  \citenamefont {Koblinger}, \citenamefont {Lassmann},\ and\ \citenamefont
  {Eisenmenger}}]{galkina1987down}%
  \BibitemOpen
  \bibfield  {author} {\bibinfo {author} {\bibfnamefont {T.~I.}\ \bibnamefont
  {Galkina}}, \bibinfo {author} {\bibfnamefont {A.~Y.}\ \bibnamefont {Blinov}},
  \bibinfo {author} {\bibfnamefont {M.}~\bibnamefont {Bonch-Osmolovskii}},
  \bibinfo {author} {\bibfnamefont {O.}~\bibnamefont {Koblinger}}, \bibinfo
  {author} {\bibfnamefont {K.}~\bibnamefont {Lassmann}}, \ and\ \bibinfo
  {author} {\bibfnamefont {W.}~\bibnamefont {Eisenmenger}},\ }\href@noop {}
  {\bibfield  {journal} {\bibinfo  {journal} {Physica Status Solidi B}\ }
  (\bibinfo {year} {1987})}\BibitemShut {NoStop}%
\bibitem [{\citenamefont {Klitsner}\ and\ \citenamefont
  {Pohl}(1987)}]{klitsner1987phonon}%
  \BibitemOpen
  \bibfield  {author} {\bibinfo {author} {\bibfnamefont {T.}~\bibnamefont
  {Klitsner}}\ and\ \bibinfo {author} {\bibfnamefont {R.}~\bibnamefont
  {Pohl}},\ }\href@noop {} {\bibfield  {journal} {\bibinfo  {journal} {Physical
  Review B}\ }\textbf {\bibinfo {volume} {36}},\ \bibinfo {pages} {6551}
  (\bibinfo {year} {1987})}\BibitemShut {NoStop}%
\bibitem [{\citenamefont {Msall}\ \emph {et~al.}(1993)\citenamefont {Msall},
  \citenamefont {Carroll}, \citenamefont {Shield},\ and\ \citenamefont
  {Wolfe}}]{msall1993observation}%
  \BibitemOpen
  \bibfield  {author} {\bibinfo {author} {\bibfnamefont {M.}~\bibnamefont
  {Msall}}, \bibinfo {author} {\bibfnamefont {M.}~\bibnamefont {Carroll}},
  \bibinfo {author} {\bibfnamefont {J.}~\bibnamefont {Shield}}, \ and\ \bibinfo
  {author} {\bibfnamefont {J.}~\bibnamefont {Wolfe}},\ }in\ \href@noop {}
  {\emph {\bibinfo {booktitle} {Phonon Scattering in Condensed Matter VII}}}\
  (\bibinfo  {publisher} {Springer},\ \bibinfo {year} {1993})\ pp.\ \bibinfo
  {pages} {116--117}\BibitemShut {NoStop}%
\bibitem [{\citenamefont {Xie}\ \emph {et~al.}(2013)\citenamefont {Xie},
  \citenamefont {Guo}, \citenamefont {Li}, \citenamefont {Yang}, \citenamefont
  {Zhang}, \citenamefont {Tang},\ and\ \citenamefont {Zhang}}]{xie2013phonon}%
  \BibitemOpen
  \bibfield  {author} {\bibinfo {author} {\bibfnamefont {G.}~\bibnamefont
  {Xie}}, \bibinfo {author} {\bibfnamefont {Y.}~\bibnamefont {Guo}}, \bibinfo
  {author} {\bibfnamefont {B.}~\bibnamefont {Li}}, \bibinfo {author}
  {\bibfnamefont {L.}~\bibnamefont {Yang}}, \bibinfo {author} {\bibfnamefont
  {K.}~\bibnamefont {Zhang}}, \bibinfo {author} {\bibfnamefont
  {M.}~\bibnamefont {Tang}}, \ and\ \bibinfo {author} {\bibfnamefont
  {G.}~\bibnamefont {Zhang}},\ }\href@noop {} {\bibfield  {journal} {\bibinfo
  {journal} {Physical Chemistry Chemical Physics}\ }\textbf {\bibinfo {volume}
  {15}},\ \bibinfo {pages} {14647} (\bibinfo {year} {2013})}\BibitemShut
  {NoStop}%
\end{thebibliography}%

\end{document}